# Measurement of the Relativistic Potential Difference across a Rotating Magnetic Dielectric Cylinder

*J.B. Hertzberg, S.R. Bickman, M.T. Hummon, D. Krause, Jr., S.K. Peck and L.R. Hunter*

*Physics Department, Amherst College, Amherst, MA 01002*

**Abstract**

According to the Special Theory of Relativity, a rotating magnetic dielectric cylinder in an axial magnetic field should exhibit a contribution to the radial electric potential that is associated with the motion of the material's magnetic dipoles. In 1913 Wilson and Wilson reported a measurement of the potential difference across a magnetic dielectric constructed from wax and steel balls. Their measurement has long been regarded as a verification of this prediction. In 1995 Pelligrini and Swift questioned the theoretical basis of the experiment. In particular, they pointed out that it is not obvious that a rotating medium may be treated as if each point in the medium is locally inertial. They calculated the effect in the rotating frame and predicted a potential different from both the Wilson's theory and experiment. Subsequent analysis of the experiment suggests that the Wilsons' experiment does not distinguish between the two predictions due to the fact that their composite steel-wax cylinder is conductive in the regions of magnetization. We report measurements of the radial voltage difference across various rotating dielectric cylinders, including a homogeneous magnetic dielectric material (YIG), to unambiguously test the competing calculations. Our results are compatible with the traditional treatment of the effect using a co-moving locally inertial reference frame, and are incompatible with predictions based on the model of Pelligrini and Swift.

**Introduction**

In 1908 Einstein and Laub[1] noted that the Special Theory of Relativity predicts that a moving magnetic dipole develops an electric dipole moment. Wilson and Wilson[2] performed an experimental test of this prediction using a rotating cylindrical magnetic dielectric material fabricated by imbedding steel balls in wax. The Wilson's found good agreement between their measurements and their theoretical predictions. Their predictions implicitly assume that each point in the cylinder can be treated as if it were in a locally inertial (LI) reference frame instantaneously co-moving with the point. In 1995, Pellegrini and Swift[3] (PS) reexamined the experiment's theoretical basis. They find that the LI treatment, when applied to the cylindrical geometry of the experiment, results in non-conservation of charge. They perform the calculation in a rotating coordinate system and conclude that in the cylindrical geometry there is no contribution to the potential associated with the moving magnetic dipoles. As a consequence, they predict a voltage different from that measured and predicted by the Wilsons.

The validity of the PS analysis has been questioned in two subsequent papers. Burrows[4] suggests that the problems in the PS treatment stem from the use of a non-orthogonal coordinate system. Weber[5] cautions that physical quantities (such as the current density) are difficult to assign in the rotating frame due to problems with the synchronization of clocks. Weber follows Einstein's prescription and uses freely falling observers to define physical quantities. The resulting calculation yields a volume charge density within the cylinder that is not present in the PS treatment. This volume charge density restores conservation of charge. Weber's analysis using the rotating coordinate system results in the same voltage difference that was predicted by the LI analysis.

Krotkov et. al. examine a steel sphere moving in a uniform magnetic field and find that any theory consistent with Maxwell's equations yields the same electric field within the sphere.[6] They argue that the result can be generalized to the Wilsons' cylinder, which consists of many steel spheres embedded in wax. Essentially, their conclusion derives from the fact that in such an assembly, the magnetization (and hence the motional electric dipole) resides only in the steel balls, where the conductivity is high. They assert that in this situation, all models produce the LI model's prediction for the potential difference. Hence, the Wilsons' experiment can not in fact distinguish between the LI and PS models. This can only be accomplished if one performs the measurement using a cylinder constructed from a uniform magnetic insulator.

In the present work, we address this limitation of the Wilsons' experiment by performing measurements with both a "homogeneous" and an "inhomogeneous" magnetic dielectric. Our "inhomogeneous" magnetic dielectric is a cylinder of steel spheres embedded in wax, similar to that used by the Wilsons, and is subject to the objections raised in Ref. 6. Our "homogeneous" cylinder is made of yttrium-iron-garnet (YIG), which is a magnetic insulator, even on the molecular scale. This "homogeneous" cylinder should be well described by its bulk properties and not subject to the objections of Ref. 6. The predictions of the PS and LI models are different and distinguishable for this cylinder.

**Theoretical Background**

We are interested in the radial voltage difference across a rotating dielectric cylinder which is magnetized along, and rotating about, its symmetry axis. There are two different theoretical models that yield different predictions for this potential difference.[2,3] The models generally have been developed for infinite cylinders with linear magnetic susceptibilities. In this and the following section we apply these models to the real cylinders and detectors used in our experiment. Expressions are derived that predict the size of the potential difference using the alternative theories and other experimentally measurable parameters.

The case of uniform linear motion provides a good starting-point for the examination of this phenomenon. Consider a slab with a dielectric constant K and permeability $\mu$ (see Fig. 1). The slab has a thickness $d$ in the y-direction, while in the x and z directions it extends far enough that edge effects may be ignored. The slab moves at constant velocity $\mathbf{v} = v\hat{\mathbf{x}}$ through a uniform lab-frame field $\mathbf{H} = H_z\hat{\mathbf{z}}$. A straightforward calculation of the lab-frame electric field within the slab yields

$$E_y = \left(1 - \tfrac{\mu_0}{\mu K}\right)\mu H_z v. \tag{1}$$

All authors agree on this result.

The controversy begins when one considers a hollow cylinder constructed of a similar magnetic dielectric material and rotating in a uniform applied field $\mathbf{H} = H_z \hat{\mathbf{z}}$. Let the cylinder have inner radius $r_1$ and outer radius $r_2$. The cylinder rotates with angular speed $\omega = 2\pi f$ about the z-axis such that a point within the cylinder has an instantaneous lab-frame velocity $\mathbf{v} = r\omega\hat{\phi}$. Wilson and Wilson argue that each point within the cylinder can be treated as if it were locally inertial (LI), and hence will experience a corresponding local electric field (see Eq. 1) in the radial direction

$$E_r^{LI} = (1 - \tfrac{\mu_0}{\mu K})\mu H_z \omega r. \tag{2}$$

We recast this equation in terms of the magnetization $\mathbf{M}$ and the local field $\mathbf{B}$ so that the result can be generalized to materials in which the magnetization is not linear. With the defining relationships $\mathbf{B} = \mu\mathbf{H}$, $\mathbf{M} = \chi_M \mathbf{H}$ and $\mu = \mu_0(1+\chi_M)$, Eq. 2 can be rewritten as

$$E_r^{LI} = (1 - \tfrac{1}{K})B_z \omega r + \tfrac{1}{K}\mu_0 M_z \omega r. \tag{3}$$

For a finite length cylinder magnetized symmetrically about its axis we have

$$\mathbf{M} = M_r(r,z)\hat{\mathbf{r}} + M_z(r,z)\hat{\mathbf{z}} \text{ and } \mathbf{B} = B_r(r,z)\hat{\mathbf{r}} + B_z(r,z)\hat{\mathbf{z}}. \tag{4}$$

Hence, the LI model predicts a radial field

$$E_r^{LI} = (1 - \tfrac{1}{K})B_z(r,z)\omega r + \tfrac{1}{K}\mu_0 M_z(r,z)\omega r. \tag{5}$$

PS argue that this approach is incorrect for a rotating cylinder and conclude that the special-relativistic polarization associated with the moving magnetic dipoles does not arise in a rotating medium. Hence, the second term of equation 5 should be dropped. The radial field in their treatment becomes

$$E_r^{PS} = (1 - \tfrac{1}{K})B_z(r,z)\omega r. \tag{6}$$

We note that for the case of the infinite cylinder with a linear permeability in a uniform field this expression is equivalent to

$$E_r^{PS} = (1 - \tfrac{1}{K})\mu H_z \omega r \tag{7}$$

as was derived in Ref. 3.

Taking the integral of Eq. (5) from $r_1$ to $r_2$ yields a radial potential difference

$$V^{LI}_{r_1 \to r_2} = (1-\tfrac{1}{K})f\int_{r_1}^{r_2} B_z(r,z)2\pi r dr + \tfrac{1}{K}f\mu_0 \int_{r_1}^{r_2} M_z(r,z)2\pi r dr. \qquad (8)$$

The integrals in the above expression represent the fluxes, $\Phi$, of **M** and **B** through the cross-section of the cylinder at a given $z$:

$$V^{LI}_{r_1 \to r_2}(z) = (1-\tfrac{1}{K})f\Phi_B(z) + \tfrac{1}{K}f\mu_0 \Phi_M(z). \qquad (9)$$

For the PS model, Eq. 6 may be similarly integrated to yield

$$V^{PS}_{r_1 \to r_2}(z) = (1-\tfrac{1}{K})f\Phi_B(z). \qquad (10)$$

Note that the LI model predicts an additional contribution to the potential difference, associated with the rotating magnetization, that is not present in the PS model.

**Accommodating Experimental Reality**

Unfortunately, the voltages we expect to observe in our apparatus differ significantly from those calculated in this idealized case. To refine the predictions, we model the system as shown in Fig. 2. The cylinder extends from $-\tfrac{L}{2}$ to $\tfrac{L}{2}$ in the $z$ dimension. A solid metal shaft fills its interior region and contacts the surface at $r_1$. The outer conductive sheath has some "thickness," extending to $r_3$. These metal surfaces contacting the dielectric cylinder effectively produce a tubular capacitor, which we assume has a uniform capacitance per unit length $\gamma$ and a total capacitance $C_{int} = \gamma L$. Two sliding brushes contact the metallic surfaces at $r_1$ and $r_3$. While the cylinder spins, the potential difference between these brushes is measured by a voltmeter in the laboratory frame. The voltmeter and wires in the lab frame have some capacitance, $C_{ext}$.

The total magnetic field **B** is due both to the applied field and the field produced by the magnetization of the cylinder. Because of the finite length of the cylinder, even for a perfectly uniform applied field, **B** possesses a small radial component. Hence the potential is a function of both $r$ and $z$.

In the experiment, the theoretical voltage difference (Eq. 9 or Eq. 10) is not registered exactly by the voltmeter, but rather is modified due to the capacitances of the system and the nonuniform **B** and **M** within the cylinder. We wish to predict for both models the voltage difference expected across our voltmeter once these effects are taken into account. To accomplish this, first consider the total charge expected to be produced on the surfaces of the spinning cylinder:

$$Q = \int_{-\frac{L}{2}}^{\frac{L}{2}} \gamma \Delta V^T_{r_1 \to r_2}(z) dz, \tag{11}$$

where $\Delta V^T_{r_1 \to r_2}(z)$ is the theoretical voltage difference given by either Eq. (9) or Eq. (10). This total charge $Q$ is redistributed between the cylinder and the external circuitry such that

$$Q = C_{ext} V_{ext} + \int_{-\frac{L}{2}}^{\frac{L}{2}} \lambda(z) dz, \tag{12}$$

where $\lambda(z)$ represents the final charge per unit length on the rotating cylinder.

To determine $V_{ext}$, we apply simple circuit theory. The sum of the voltages around any closed conducting loop is zero (Kirchhoff's Voltage law). Referring to figure 2, consider a "loop" from the axis extending radially across the cylinder at the position of the outer brush, through the wire to the external capacitance and back through the second brush to the axis again. This loop yields the relation

$$\Delta V_{0 \to r}(z_1) + \frac{q_{ext}}{C_{ext}} - \Delta V_{r_2 \to r_3}(0) - \frac{\lambda(0)}{\gamma} - \Delta V_{0 \to r_1}(0) = 0. \tag{13}$$

Note that we have used the fact that the rotation axis remains an equipotential in the laboratory frame.

Next, consider a loop within the dielectric cylinder, spanning r from the inner to the outer radius and an infinitesimal distance along z. This yields the relation

$$\frac{d}{dz}(V(r_2, z) - V(r_1, z)) - \frac{d}{dz}\left(\frac{\lambda(z)}{\gamma}\right) = 0. \tag{14}$$

Eq. (14) may be integrated from $z = 0$ to $z'$ to yield an expression for $\lambda(z')$. Combining the result with Eq. (13) to eliminate $\frac{\lambda(0)}{\gamma}$ yields

$$\lambda(z') = \gamma\left(\Delta V_{0 \to r}(z_1) - \Delta V_{r_2 \to r_3}(0) - \Delta V_{0 \to r_1}(0) + \frac{q_{ext}}{C_{ext}} + \Delta V_{0 \to z'}(r_2) - \Delta V_{0 \to z'}(r_1)\right). \tag{15}$$

Replacing $z'$ in Eq. (15) with $z$, inserting the expression for $\lambda(z)$ into Eq. 12, and evaluating the integral yields

$$q_{ext} = \frac{C_{ext}}{C_{ext} + C_{int}} \left( \begin{array}{l} Q + C_{int}\left(\Delta V_{r_2 \to r_3}(0) + \Delta V_{0 \to r_1}(0) - \Delta V_{0 \to r_1}(z_1)\right) \\ + \int_{-\frac{L}{2}}^{\frac{L}{2}} \gamma(\Delta V_{0 \to z}(r_1) - \Delta V_{0 \to z}(r_2))dz \end{array} \right). \quad (16)$$

We restate the above expression as

$$q_{ext} = \frac{C_{ext}}{C_{ext} + C_{int}} \left( \begin{array}{l} C_{int}\left(\Delta V_{r_2 \to r_3}(0) + \Delta V_{0 \to r_1}(0) - \Delta V_{0 \to r_1}(z_1)\right) \\ + C_{int}\int_{-\frac{L}{2}}^{\frac{L}{2}} \frac{1}{L}\left(\Delta V_{r_1 \to r_2}^T(z) + \Delta V_{0 \to z}(r_1) - \Delta V_{0 \to z}(r_2)\right)dz \end{array} \right). \quad (17)$$

where $\Delta V_{r_1 \to r_2}^T(z)$ is again given by either Eq. (9) or Eq. (10). Note that the potential differences $\Delta V_{0 \to z}(r_1)$ and $\Delta V_{0 \to z}(r_2)$ are a consequence of the Lorentz force acting along the conducting shaft and outer sheath respectively. If the dielectric material between $r_1$ and $r_2$ were conducting rather than insulating, but retained its magnetic characteristics, these two potential differences would remain unchanged. Hence, we may replace $\Delta V_{0 \to z}(r_1)$ by $\left(f\Phi_{B,0 \to r_1}(z) - f\Phi_{B,0 \to r_1}(0)\right)$ and $\Delta V_{0 \to z}(r_2)$ by $\left(f\Phi_{B,0 \to r_2}(z) - f\Phi_{B,0 \to r_2}(0)\right)$. The other potential differences may also be expressed in terms of the magnetic fluxes, for example $\Delta V_{0 \to r_1}(z_1)$ by $\left(f\Phi_{B,0 \to r_1}(z_1)\right)$. Employing these substitutions, and using Eq. (9) for $\Delta V_{r_1 \to r_2}^T(z)$ yields

$$q_{ext} = f \frac{C_{int}C_{ext}}{C_{ext} + C_{int}} \left( \begin{array}{l} \Phi_{B,0 \to r_1}(0) + \Phi_{B,r_1 \to r_2}(0) + \Phi_{B,r_2 \to r_3}(0) - \Phi_{B,0 \to r_1}(z_1) \\ + \mu_0 \frac{1}{K}\langle\Phi_{M,r_1 \to r_2}\rangle - \frac{1}{K}\langle\Phi_{B,r_1 \to r_2}\rangle \end{array} \right), \quad (18)$$

where the brackets $\langle\,\rangle$ denote the average of a quantity over the length of the cylinder.

We simplify the expression further by noting that the first four terms inside the parentheses represent the voltage difference, $V_{short}$ that would be measured if the spinning magnetized cylinder is "shorted" so that charge flows freely between the inner and outer surfaces. The voltage measured by the lab-frame voltmeter divided by the rotation frequency can then be expressed as

$$V_{ext}^{LI}/f = \frac{C_{int}}{C_{ext} + C_{int}}\left(\tfrac{1}{f}V_{short} - \tfrac{1}{K}\langle\Phi_{B,r_1 \to r_2}\rangle + \tfrac{\mu_0}{K}\langle\Phi_{M,r_1 \to r_2}\rangle\right). \quad (19)$$

A similar set of substitutions employing Eq. (10) for $\Delta V_{r_1 \to r_2}^T(z)$ yields

$$V_{ext}^{PS}/f = \frac{C_{int}}{C_{ext} + C_{int}}\left(\tfrac{1}{f}V_{short} - \tfrac{1}{K}\langle\Phi_{B,r_1 \to r_2}\rangle\right). \quad (20)$$

This completes our derivation. Eqs. (19) and (20) predict respectively the values of the voltages we should measure according to the LI and PS models. The predictions are expressed in terms of system parameters that can be measured or calculated. The expressions remain valid if the position of the brushes is changed, but the value of $V_{short}$ does depend upon the placement of the two brushes. The potential differences in Eqs. (19) and (20) arise directly from the predictions of the radial potential in a rotating magnetic dielectric outlined in Refs. 2 and 3 respectively.

These expressions motivate choices of favorable experimental parameters. It is evident that Eqs. (19) and (20) are most clearly distinguishable if the dielectric constant of the material is small and the induced magnetization is large. It is also evident that the largest possible measured potential is obtained when the voltmeter's capacitance is negligible compared to the capacitance of the cylinder. Small nonzero conductivity in the capacitors imposes another experimental restriction because each capacitor tends to "short" with a characteristic RC time constant (see Table I.). If the external capacitance is "leaky," the measured signal decays toward zero, while if the internal capacitance is "leaky," it rises towards the shorted value, making the predictions of Eqs. (19) and (20) indistinguishable. To minimize such considerations, measurements are made in a time much shorter than RC.

**Measurement of the Parameters Required to Deduce the Expected Voltages**

In order to compare experimental measurements effectively with Eqs. (19) and (20), it is necessary to know the "shorted" voltage, the average fluxes of **M** and **B** within the cylinder, and the capacitance ratio $\frac{C_{int}}{C_{ext} + C_{int}}$.

The measurement of $V_{short}$ is of central importance to the success of the experiment. Fortunately, it is straightforward. Aluminum foil is molded over the upper end of the cylinder, establishing good electrical contact between the inner and outer conductive surfaces. In this configuration, the low-impedance potential difference between the brushes is easily measured with high precision. The cylinder works as a homopolar generator and the resulting potential difference is independent of where the short between the inner and outer cylinder surfaces is made. Knowing this potential significantly relaxes the accuracy with which the other parameters of our system must be known. We essentially only need to model the small departures from this "shorted" voltage, rather than requiring a high precision prediction for the full voltage.

The values of $\langle \Phi_{M, r_1 \to r_2} \rangle$ and $\langle \Phi_{B, r_1 \to r_2} \rangle$ are deduced from a numerical simulation of the fields within the experimental cylinder. We assume **M** is uniform throughout the cylinder and that the external applied magnetic field is known. We employ the relation $\mathbf{B} = \mu_0 (\mathbf{H} + \mathbf{M})$ and Maxwell's equation $\nabla \cdot \mathbf{B} = 0$ to solve numerically for **H**, and therefore **B**, at all positions within and near the cylinder. Resultant values for **B** are integrated across a cross-section of the cylinder to calculate values for $\Phi_{B, 0 \to r_3}(z)$. Measurements of $\Phi_{B, 0 \to r_3}(z)$ are made inductively by wrapping a search-coil of fine wire around the cylinder and integrating the voltage change across the coil that arises when the applied magnetic field is reversed. The materials used in our magnetic cylinders are highly permeable, allowing **M** at all points within the cylinder to be reversed by the reversal of the applied magnetic field. (A similar modulation technique is used in the actual measurements of the radial potential to isolate the signal and improve the signal-to-noise ratio). Fig. 3 compares values of $\Phi_{B, 0 \to r_3}(z)$ measured using this technique with values calculated on the assumption that **M** is uniform within the cylinder. The close agreement between the two suggests that this assumption is reasonable and sufficiently accurate for our purposes. Values of $\langle \Phi_{M, r_1 \to r_2} \rangle$ and $\langle \Phi_{B, r_1 \to r_2} \rangle$ are then calculated from this numerical model.

The value of $\dfrac{C_{int}}{C_{ext} + C_{int}}$ is measured by connecting the cylinder and voltmeter in series across an isolated, low impedance voltage source. The sign of the applied voltage is reversed with precisely the same cadence as the reversal of the applied magnetic field in the experiment (see "Design of the Apparatus" below). Analysis of the simple circuit reveals that the factor $\dfrac{C_{int}}{C_{ext} + C_{int}}$ is equal to the ratio of the change in the measured voltage to the change in the source voltage. This factor is measured with each data set. The similarity of the measurements ensures that any possible leakage effects (which are anticipated to be less than 0.3% for the YIG) associated with the finite resistance of the capacitors are accounted for correctly.

**Design of the Apparatus**

Our experiment is conceptually identical and similar in design to the Wilsons' experiment. The magnetizable dielectric medium rotates about a vertical axis in the presence of a very uniform vertical magnetic field. The applied field induces a magnetization, **M**, in the material. The radial potential developed in the spinning dielectric is measured as a function of rotational frequency and the results are compared with the theoretical predictions Eqs. (19) and (20).

A schematic of our apparatus is shown in Fig. 4. A trimmed magnetic coil is wrapped on a large brass cylinder to create the uniform applied field. Conductive caps over the top and bottom of this cylinder create a Faraday cage that isolates the cylinder from external electric fields. An encoded motor placed below the apparatus spins the cylinder via a non-magnetic stainless-steel shaft. The cylinder is electrically isolated from the drive shaft and motor by a polycarbonate insert within the inner brass cylinder. Electrical brushes made of "cophite", a copper-graphite mixture, contact the spinning cylinder on the inner and outer surfaces. An ultra-low bias current instrument amplifier (Burr Brown, model 1NA116) measures the voltage difference between the brushes. Both input leads of this amplifier have an impedance with respect to ground of $2 \times 10^{11}$ $\Omega$. The amplifier transforms the high impedance voltage difference to a low impedance voltage that is then digitized by a Keithley 2000 voltmeter. A precision, fan-cooled, 0.332 $\Omega$ resistor is placed in series with a compensated magnetic solenoid, designed to produce a highly uniform magnetic field. A Fluke digital voltmeter monitors the voltage across this resistor, and hence (using the measured calibration) the applied magnetic field strength. The digitized voltages from the two commercial voltmeters as well as the rotation frequency of the cylinder, as determined from an encoded signal from the drive motor, are stored for analysis by the experimental running program.

Two different magnetic dielectrics have been tested in our apparatus. The first is a composite of 1/8" diameter steel balls embedded in wax, very similar to the cylinder constructed by the Wilsons. This cylinder is fabricated by first dipping the chrome-steel balls in paraflint H1 wax (melting temperature (MT) = 100 C). A layer of these balls is placed between the inner brass shaft and the thin outer brass sheath. A small amount of Triacontane wax (MT = 67 C) is then poured in to complete the layer and hold the balls in place before adding an additional layer of balls. The process is repeated until the entire length of the cylinder is filled with the steel-wax composite. The average dielectric constant for this cylinder is measured to be K = 2.85 (20) while its average magnetic permeability is about µ = 1.80. Our values for µ and K are both smaller than those of the Wilsons' steel-wax composite cylinder. We suspect that the steel balls in the Wilsons' cylinder were more efficiently packed, since this would account for their higher values for µ and K. Our cylinder has a volume percentage of just 22% steel.

The material used in our other magnetic dielectric cylinder is Yttrium Iron Garnet (YIG). This cylinder, unlike the Wilsons', is both magnetically and electrically homogeneous. The dielectric constant of the YIG is $\kappa$ = 15 (1). Our magnetic modeling of the cylinder suggests that the YIG's magnetization is very near its saturated value (about 1860 G) at the applied magnetic fields used in the experiment. To insure good electrical and mechanical contact with the cylinder, silver epoxy is used to affix the inner brass shaft and outer brass sheath to the YIG cylinder. The YIG is cooled with a large fan located outside of the solenoid as well as by blowing bench air directly on to the rotating cylinder. The cylinder's temperature is carefully monitored since the magnetization of the YIG varies with temperature. The cylinder temperature is never allowed to change by more than two degrees Centigrade during data collection.

A non-magnetic dielectric cylinder is also tested. In such a cylinder, **M** is zero, and the predictions of Eqs. (19) and (20) for the radial potential are identical. This cylinder was constructed of a nylon tube sandwiched between a brass shaft and a thin-walled brass tube. Some of the parameters associated with our three cylinders are listed in Table I.

One of the most significant experimental problems is that the voltage of interest is small compared to the electrical noise created by spinning the cylinder. Furthermore, the parameters of the YIG are such that the difference between the predicted voltages

[Eqs. (19) and (20)] is only about 6%. A measurement with an accuracy of about 1% is hence required to clearly distinguish between the two predictions. To achieve this level of precision requires a large number of measurements. To eliminate voltage offsets in our system, the direction of the current in the solenoid (and hence the applied magnetic field) is reversed every 200 ms by a relay, thereby reversing both **M** and **B** and changing the sign of the signal. The voltage difference is recorded at four points, when the current is +, -, -, and finally +. After taking into account the sign, the four measurements are averaged to form a single data point. This sequence eliminates linear drifts in the voltage. Fifty such points are taken at several rotation frequencies for each configuration of the cylinders. We assign a positive or negative sign to the frequency depending upon the sense of rotation of the cylinder. The slope obtained from a 2 parameter linear regression of the measured voltage differences versus the frequency is then compared to the theoretical predictions. A non-zero intercept for this line indicates the presence of an induced potential produced by the reversal of the magnetic field. Careful placement of the brushes and wires maintains this effect at an acceptably small level.

**Results**

Data are collected in the above manner for each of our cylinders. The magnetic cylinders are examined at two different applied fields in order to investigate possible effects associated with incomplete saturation of the materials. In addition, for the YIG cylinder, data are collected at two different axial positions of the outer brush to test the consistency of the theoretical formalism developed above. One set of YIG data is taken with the voltmeter located outside of the electrostatic shields. In this set, the signal is sent to the voltmeter on two triaxial cables, which significantly increase $C_{ext}$ and the size of the capacitive correction.

A typical set of data is shown in Fig. 5. The results obtained from each of the eight configurations investigated are summarized in Table II. Mechanical resonances in the apparatus can result in unusually high noise at particular rotation frequencies. In addition to increasing the noise, such data tends to have a slightly lower average value of V/f than that obtained from quieter data. This is consistent with the postulate that the excess noise is associated with intermittence in the brush contacts due to the mechanical resonance, which also results in a decrease in the size of the measured voltage difference. Data collected near such resonance frequencies, as determined by their anomalously large standard error, have not been included in the analysis.

All of the data in Table II are reasonably consistent with the LI predictions of Eq. (19). The agreement between the data collected with different positions of the brushes and of the voltmeter lend confidence to our electrical modeling of the cylinder and our ability to make the capacitive correction.

The data taken with the magnetic cylinders are clearly not consistent with the PS predictions of Eq. (20). The authors of Reference 6 suggest that for a cylinder composed of steel balls in wax, one in fact does not expect Eq. (20) to hold, but that Eq. (19) will describe the anticipated voltage, independent of which theoretical model is invoked. Hence, our steel and wax results, like those of the original Wilson and Wilson experiment, cannot definitively distinguish between the alternative theories. However, our results for the YIG cylinder contradict unambiguously the PS predictions from Eq. (20), and hence the formalism developed in Ref. 3 by Pellegrini and Swift.

We wish to thank the Pacific Ceramics Company for the fabrication and donation of the YIG cylinder and Phillip Grant for important technical support. We thank Prof. Robert H. Romer, Prof. Robert V. Krotkov, Prof. Kannan Jagannathan and Dr. Gerald N. Pellegrini for helpful discussions. This work was supported by NSF RUI grants PHY-9722611 and PHY-9987863. JBH and SRB were supported by NSF REU supplemental grants. MTH and SKP would like to thank Amherst College for partial support during this work.

|  | Nylon | YIG | Wax and Steel Ball |
|---|---|---|---|
| $r_1$ (cm) | 1.267 (5) | 1.266 (5) | 1.272 (5) |
| $r_2$ (cm) | 1.811 (5) | 1.803 (5) | 1.771 (5) |
| $r_3$ (cm) | 1.910 (5) | 1.905 (5) | 1.905 (5) |
| Capacitance (pF) | 58 (2) | 214.0 (2) | 81 (1) |
| R ($\Omega$) | 2.2 (2) x $10^{12}$ | 2.62 (3) x $10^{10}$ | 1.20 (6) x $10^{12}$ |
| RC (sec) | 128 (13) | 5.61 (6) | 97 (5) |
| K | 4.4 (4) | 15 (1) | 2.85 (20) |

Table I. Characteristics of the various cylinders. All of the dielectric cylinders are 3.5 inches long while the central brass shafts are 4.75 inches long.

| Cylinder (r$_3$ brush position) | Field (G) | Shorted Signal (μV/Hz) | $\langle\Phi_B\rangle$ (μT-m$^2$) | $\langle\Phi_M\rangle\mu_0$ (μT-m$^2$) | Cap. Correction | Prediction (LI) (μV/Hz) | Prediction (PS) (μV/Hz) | Measured Signal (μV/Hz) |
|---|---|---|---|---|---|---|---|---|
| Nylon | 262 | 17.22 (1) | 13.4 (2) | 0 | 0.8823 (12) | 12.5 (2) | 12.5 (2) | 12.2 (3) |
| Steel & Wax | 221 | 21.66 (5) | 19.4 (7) | 9.1 (8) | 0.8029 (84) | 14.5 (4) | 11.9 (5) | 14.57 (25) |
| Steel & Wax | 262 | 26.01 (2) | 23.0 (6) | 10.8 (6) | 0.8150 (58) | 17.7 (4) | 14.6 (5) | 16.80 (28) |
| YIG (center) | 221 | 92.19 (17) | 96 (4) | 97 (6) | 0.9445 (26) | 87.1 (5) | 81.0 (6) | 86.65 (12) |
| YIG (center) | 262 | 95.20 (4) | 98 (4) | 97 (6) | 0.9375 (61) | 89.2 (7) | 83.1 (7) | 89.21 (5) |
| YIG (end) | 221 | 80.75 (10) | 96 (4) | 97 (6) | 0.9511 (15) | 76.9 (5) | 70.7 (5) | 75.14 (27) |
| YIG (end) | 262 | 85.19 (25) | 98 (4) | 97 (6) | 0.9500 (15) | 80.8 (5) | 74.7 (6) | 80.21 (9) |
| YIG [ext] | 221 | 92.48 (10) | 96 (4) | 97 (6) | 0.7786 (26) | 72.1 (4) | 67.0 (5) | 72.05 (20) |

Table II. Comparison of measured voltages to the predictions of the locally inertial (LI) and Pelligrini and Swift (PS) theories. The file labeled [ext] was taken with the outer brush centered and the voltmeter located outside of the electrostatic shields.

Figure Captions

Figure 1: A schematic of a moving, magnetically permeable, dielectric slab.

Figure 2: A schematic of the electrical model used to infer the expected voltage difference measured by our voltmeter (V). The cylinder is brass in the regions with $r<r_1$ and $r_2< r<r_3$. The region between $r_1$ and $r_2$ is filled with magnetic dielectric. $C_{ext}$ denotes the external capacitance of the voltmeter and wires.

Figure 3: The flux of B through the YIG cylinder at an applied magnetic field of 262 G. The crosses are the measured values. The solid curve is the result our simulation which assumes a uniform magnetization of the sample. The dashed line is a 4'th order polynomial fit to the measurements at $r_1$ and $r_3$.

Figure 4: A schematic representation of the experimental apparatus.

Figure 5: A typical data set along with the two predictions. The data shown is for the YIG cylinder with an applied magnetic field 262 G and the outer brush axially centered.

Figure 1

J.B. Hertzberg et. al.

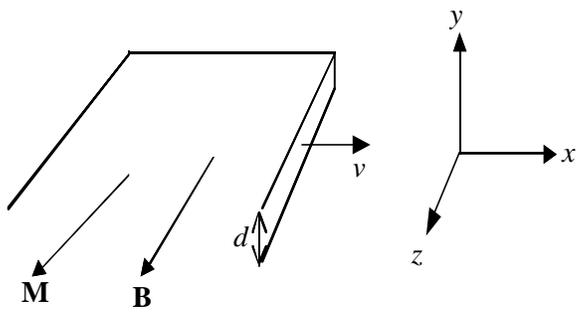

Figure 2

J.B. Hertzberg et. al.

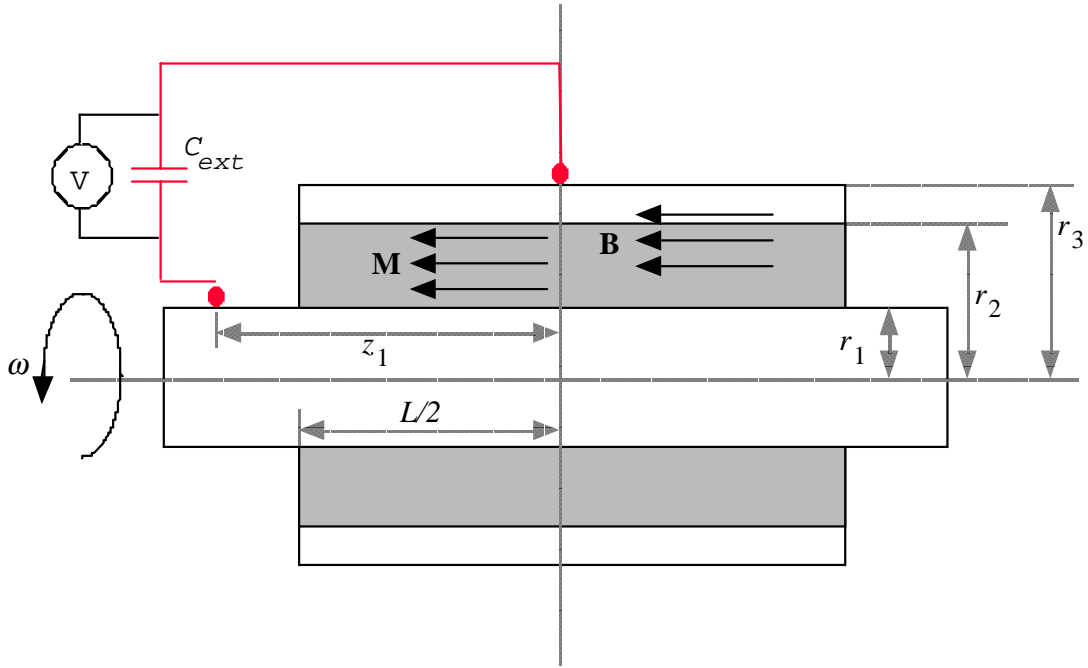

Figure 3.

J.B. Hertzberg et. al.

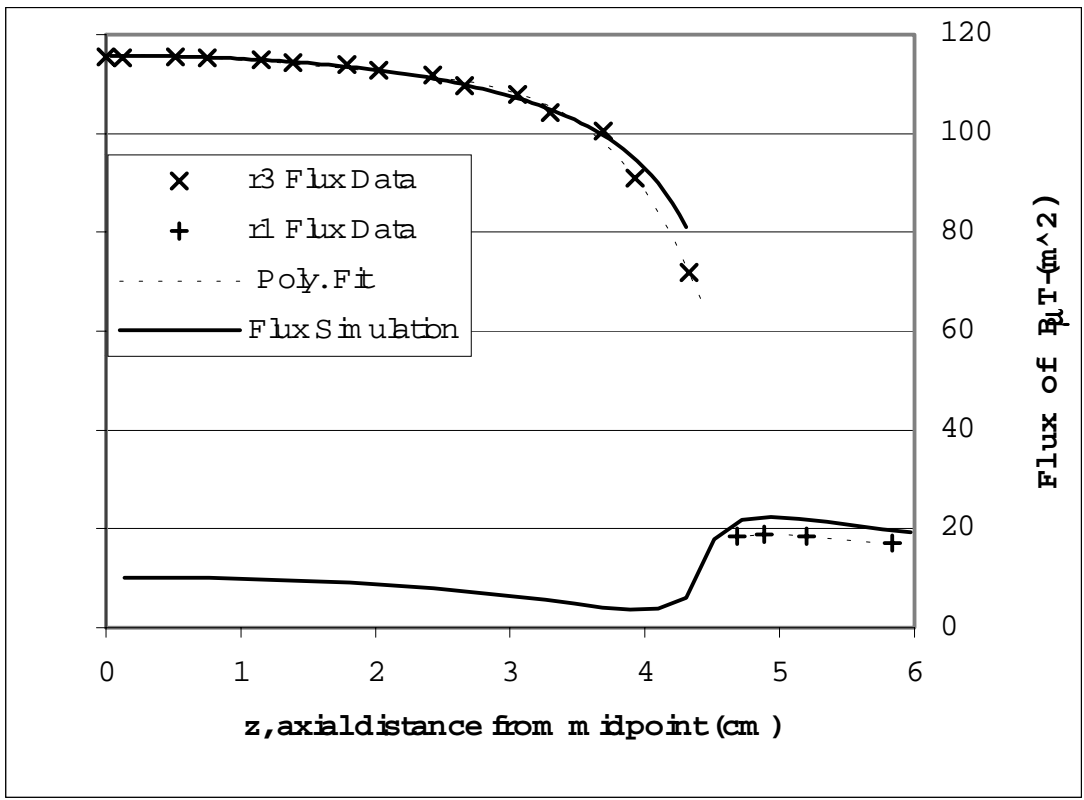

Figure 4

J.B. Hertzberg et. al.

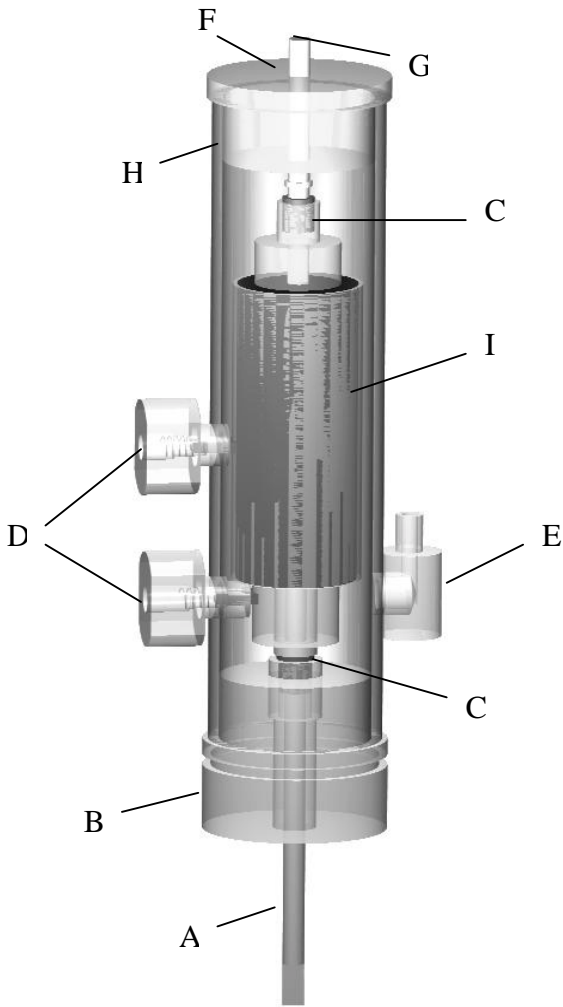

| A Stainless steel shaft | F Brass Cap |
| B Brass base | G Stainless screw to stabilize cylinder |
| C Phosphor bronze bearing | H Polycarbonate tube |
| D Brush mount | I Cylinder |
| E Air cooling nozzle | |

Figure 5

J.B. Hertzberg et. al.

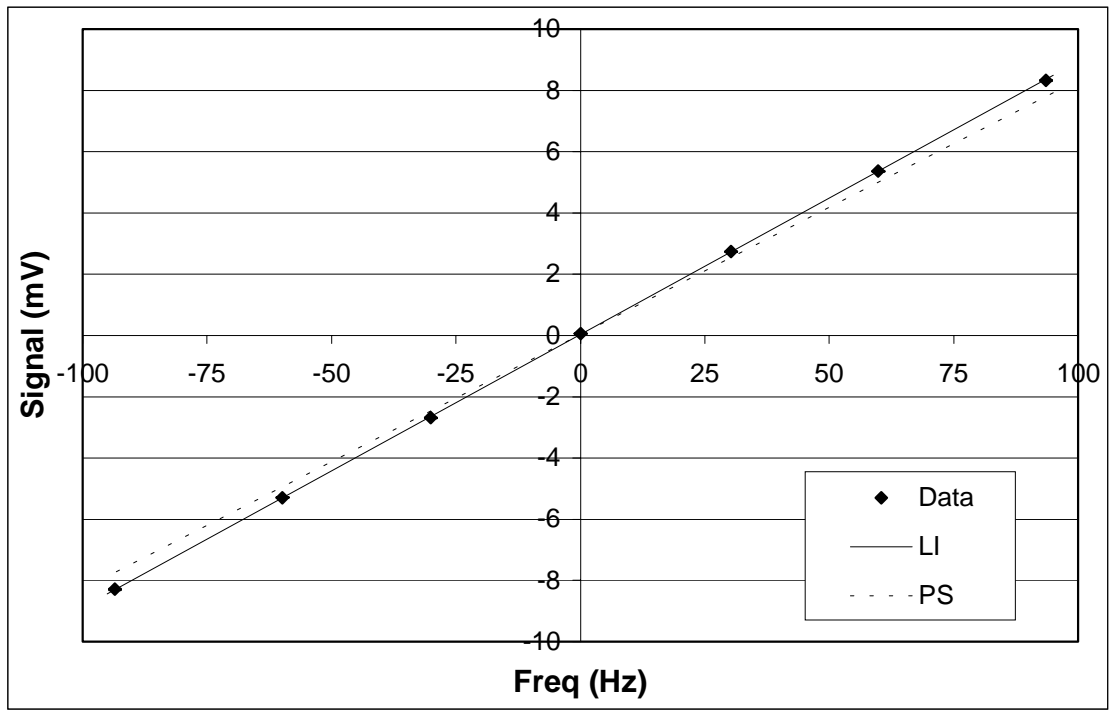